\documentclass[runningheads,a4paper]{llncs}
\usepackage{amssymb}
\usepackage{graphicx}
\usepackage{diagbox}
\usepackage{listings}
\usepackage{amsmath}
\usepackage{url}
\urldef{\mailsa}\path|{alfred.hofmann, ursula.barth, ingrid.haas, frank.holzwarth,|
\urldef{\mailsb}\path|anna.kramer, leonie.kunz, christine.reiss, nicole.sator,|
\urldef{\mailsc}\path|erika.siebert-cole, peter.strasser, lncs}@springer.com|
\newcommand{\keywords}[1]{\par\addvspace\baselineskip
\noindent\keywordname\enspace\ignorespaces#1}

\graphicspath{{figures/}}
\usepackage{subfigure}
\usepackage{algorithm }
\usepackage{algorithmic}

\begin{document}

\mainmatter  

\title{Peeking the Impact of Points of Interests on Didi}

\titlerunning{Peeking the Impact of Points of Interests on Didi}

%
%

\author{Yonghong Tian, Zeyu Li, Zhiwei Xu, Xuying Meng, and Bing Zheng}

\authorrunning{Yonghong Tian\and Zeyu Li\and Zhiwei Xu\and Xuying Meng\and Bing Zheng}

\institute{}

%
%

\toctitle{Lecture Notes in Computer Science}
\tocauthor{Authors' Instructions}
\maketitle

\begin{abstract}
Recently, the online car-hailing service, Didi, has emerged as a leader in
the sharing economy. Used by passengers and drivers extensive, it becomes
increasingly important for the car-hailing service providers to minimize the
waiting time of passengers and optimize the vehicle utilization, thus to
improve the overall user experience. Therefore, the supply-demand estimation
is an indispensable ingredient of an efficient online car-hailing service.
To improve the accuracy of the estimation
results, we analyze the implicit relationships between the points of
Interest (POI) and the supply-demand gap in this paper. The different categories
of POIs have positive or negative effects on the estimation, we propose a
POI selection scheme and incorporate it into XGBoost \cite{IEEEhowto:Tianqi} to achieve
more accurate estimation results. Our experiment demonstrates our method provides
more accurate estimation results and more stable estimation results than
the existing methods.
\end{abstract}

\section{Introduction}
Recently, there has been an explosion of service that facilitate the
so-called ``sharing economy''. In terms of transportation, several car-hailing
mobile apps have gained great popularity all over the world, such as Uber,
Didi, and Lyft. Online car-hailing apps/platforms have emerged as a novel
and popular means to provide on-demand transportation service via mobile
apps. Comparing with the traditional transportation such as the subways and
buses, the online car-hailing service is much more convenient and flexible
for the passengers. Furthermore, by incentivizing private cars owners to
provide car-hailing services, it promotes the sharing economy and enlarges
the transportation capacities of the cities. Large number of passengers are
served and volume of car-hailing orders are generated routinely every day.

As a large number of drivers and passengers use the service, several issues
arise: Occasionally, some drivers experience a hard time to get any request since few people nearby call the rides; At the same time, it is very difficult for some passengers to get the ride, in bad weather or rush hours, because the demand in the surrounding areas significantly exceeds the supply. Hence, it is a very important yet challenging task for the service providers to schedule the drivers in order to minimize the waiting time of passengers and maximize the driver utilization. One of the most important ingredient of an effective driver scheduler is the supply-demand estimation.

There are many factors that affect car-hailing supply-demand. According to
\cite{IEEEhowto:Le Chen}, weather and time and traffic condition can directly affect the
car-hailing supply-demand gap. Moreover, some obliterate factors also have
an important impact on the car-hailing supply-demand. If we can analyze
these potential factors, we can use them to establish an estimation model
According to daily observation the infrastructures of a block will affect
the car-hailing supply-demand. People usually like to go to downtown,
because better infrastructure there. Inspired by this essential
reason, we take Points of Interest (POI) into consideration.

However, considering limited information of POIs(only a numerical identifier),
it is not a trivial work to choose POIs adaptively from the great number of the
different kinds of POIs in the analysis of the supply-demand gap. In this paper, we propose a method called PPCE-SDG to choose a few significant POIs. We firstly use K-MEANS algorithm to cluster blocks based on supply-demand gap. According to clustering results, we count POIs of each cluster and get a data vector of POIs. Finally, we act PCA(Principal Component Analysis)
algorithm on the data vector to extract out the POIs.

In order to understand the impact of POI on car-hailing supply-demand, we
use an end-to-end framework called XGBoost to
conduct experiments. We use the public data-set released by Didi in the
Di-tech supply-demand estimation competition. The data contains the
car-hailing orders from Didi over more than 7 weeks of 58 square blocks in
Hangzhou, China.The order data-set consists of 11,467,117 orders. Auxiliary
information include POI information(distribution of POI in each block) and
weather conditions (temperature, PM 2.5) and traffic conditions (total
amount of road segments under different congestion levels in each area).

Additionally, our data enables us to identify impact details of POIs. Based
on our analysis of POIs, we make the following observations:

\begin{itemize}
\item POIs have relation with the car-hailing supply-demand. In the experiment, the estimation accuracy of models based on POIs is extremely different from the accuracy of the model without POIs. From the point of the car-hailing supply-demand, blocks with similar POIs have similar situation about using the car-hailing, on the contrary, blocks with non-similar POIs are extremely different in using the online car-hailing.
\item Different POIs have different impact. When the same number of POIs from different types are selected into the models, the accuracy of the estimation is incredibly different. With the increase of the number of POIs from different types selected into the models, the accuracy firstly rises and then falls, which shows different categories of POIs have different effects on estimation accuracy. Therefore, we should focus to choose POIs to achieve higher estimation accuracy.
\item A few POIs play the decisive roles. We use PCA algorithm to choose the primary POIs. After verification, the effect of a few important POIs almost represent the effect of all POIs. In the experiment, four most important POIs even represent 90{\%} of POIs in terms of improving the estimation accuracy.
\end{itemize}
The benefit of our analysis about impact of POIs on car-hailing supply-demand is
obvious. If one could accurately predict/estimate how many
passengers need the ride service in a certain area and how many close-by
drivers are available, it is possible to balance the supply-demands in
advance by dispatching the cars, dynamically adjusting the price, or
recommending popular pick-up locations to some drivers.

We begin in {\S}2 by introducing background and related work. In {\S}3, we
analyze the impact of POIs on predicting, and elaborate the details of
choosing significant POIs. In {\S}4, we describes the experimental
information and model and synthetically analyze the experimental results,
followed by conclusion in {\S}5.

\section{Related Work}

Most of existing related work is not specific, such as \cite{IEEEhowto:Saadi} which only
compared machine learning approaches for predicting supply-demand gap for
ride-hailing services. There are also some works which simply detect the
hot-spots of passengers like \cite{IEEEhowto:Han Wen} and \cite{IEEEhowto:Qiu Yan}. The same problem also appears in
\cite{IEEEhowto:Santi}, which only explored the vacant car position.

Though there are lots of other features can be used by this task, such as
weather information, traffic condition, the point of interest(POI) and so
on. For example, the bad weather will force more people to choose an online
car, and one location with more POIs like shopping malls and restaurants
will have a higher requirement of the online car than other locations. On
the contrary, traffic jam would decrease people's interest of taking the
online car even if they need. Even \cite{IEEEhowto:Xiao Zhang,IEEEhowto:Dong Wang,IEEEhowto:Xiqun,IEEEhowto:Weifeng}, design an excellent
model for online car driver's itinerary recommendations and route planning,
but they still ignore the specific contribution of POIs. In research of
Uber's surge price algorithm, although \cite{IEEEhowto:Le Chen} observed that the supply-demand
gap has relationship with POIs, they neither demonstrated this relationship
nor made a thorough inquiry.

The estimation accuracy of the existing methods suffer from their
insufficient information without the information of the points of interests
. Among these information,  location information is indispensable to predict
the taxi requirement, including POIs, traffics of the neighboring blocks.
POIs are the genetic internal factor of the gap between the car-hailing
supply and demand. Different types of POIs have distinct impact on the taxi
requirement, e.g, shopping malls have positive impact on the taxi
requirement. As the external factor of the supply-demand gap the traffics of
the neighboring blocks have not explicitly inferred in the existing work as
well as POIs, the congestion degree of the nearby roads have associated the
traffics of the neighboring blocks with the estimation results. To achieve
more accurate estimation results, we therefore analyze the impact of the
POIs extensively in the following section and leverage the analysis
conclusion to refine the XGBoost framework.

\section{Analysis}

In this Section, We firstly describe the data-set of Didi and briefly introduce the basic situation of POIs. In Section 3.2, we demonstrate the relationship between POIs and car-hailing supply-demand. Then, we illustrate the contributions of POIs are distinct for estimation in Section 3.3. Finally, we propose a selection method called PPCE-SDG in Section 3.4.

\subsection{Data Description}

In our experiment, we use the public data-set released by Didi in the Di-tech supply-demand prediction competition.The order data-set contains the car-hailing orders from Didi over more than 7 weeks of 58 square blocks in Hangzhou, China. Auxiliary information include POI information(distribution of POI in each block) and weather conditions (temperature, PM 2.5) and traffic conditions (total amount of road segments under different congestion levels in each area). The weather information and traffic information could be added directly to the XGBoost to predict. However, the POI information is semi-structured, so we need to do data preprocessing.

It is shown that there are 173 categories of POIs from the POI data-set. All of POIs are represented by cascading numbers so that we don’t know what they are specific. Moreover, in each block, the distribution of POIs is random and the number of POIs are extremely different. In order to further analyze the POIs, we number the 173 categories of POIs from 1{\#} to 173{\#}and separately count the types and amount of POIs in each block.


For all of experiments in this paper, the training data is from 23th Feb to 17th March (24 days in total). To construct the training set, for each block in each training day, we generate one training item every 10 minutes from 0:00 to 24:00. Thus, we have 58(blocks) *24(days)* 144(items) = 200,448 training items in total. The test data is from 25th March to 31th April (28 days in total). To construct the testing set, for each block in each training day, we generate one testing item every 10 minutes from 0:00 to 24:00. Thus, we have 58(blocks) *7(days)* 144(items) = 58,464 testing items in total.

To analyze and verify, we use integrated learning framework-XGBoost as a experimental model. XGBoost is a machine learning algorithm applied to regression, which is based on the Gradient~Boosting~Decision~Tree (GBDT). XGBoost has advantages of parallelization and processing of high efficiency, so it is very suitable for dealing with big data problems. In this paper, we do each experiment for 50 times and count the estimation results after each time. To make our experiments more convincing, our final results is the average of the results in the best 10 times.

\subsection{POIs and Car-hailing Supply-Demand}
According to the existing observation, time and weather and traffic will
directly affect the car-hailing supply-demand. For example, in the morning,
the demand tends to surge in the residential areas whereas the demand in the
evening usually tends to surge in the business areas . Moreover, people need
more car-hailing in the bad weather. However, we surmise other significant factors are
also neglected in estimating the supply-demand gap, such as POIs.

In order to demonstrate our conjecture, we randomly select different number of POI to construct 4 POI collections. The number of POI corresponding to each collection are respectively 5, 10. The estimation results with these collections of POI are compared with estimation results without POI(0 POI). We calculate minimum value, average value, maximum value and standard deviation of estimation accuracy as evaluation metrics. The results are shown in the table \ref{tab3-1}:

\begin{table}[!h]
  \centering
  \caption{estimation accuracy with different number of POIs}\label{tab3-1}
  \tabcolsep = 6pt
  \begin{tabular}{|c|c|c|c|c|c|c|}
  \hline
\diagbox{Metric}{POI Number}&
0&
5&
10 \\
\hline
Minimum Value&
0.59032&
0.63359&
0.63549 \\
\hline
Average Value&
0.58839&
0.63654&
0.63787 \\
\hline
Maximum Value&
0.59261&
0.63932&
0.64019 \\
\hline
Standard Deviation&
0.00624&
0.00287&
0.00235 \\
\hline
  \end{tabular}
\end{table}

From table \ref{tab3-1}, the average values with POIs are obviously more
higher than the average value without POIs, the standard deviations with POIs are more lower than the standard deviations without POIs, which show that the addition of POIs is beneficial to the car-hailing supply-demand estimation. However, this is
only a superficial phenomenon. We will continue to analyze the deep relation between
POIs and car-hailing supply-demand.

The city is divided into 58 blocks. The situation of car-hailing
supply-demand vary widely between the blocks.At first, we use training data to count the average of the supply-demand gap of each 10 minutes in each block. The data vector of supply-demand gap is 144
dimensions(one day has 144 time slices). Moreover, considering to the
differences of supply-demand gap between the workdays and holidays, we
respectively establish 144 dimensions data vector of workdays and holidays
We combine the two data vectors and get a 288 dimensions(144*2 dimensions)
data vector. Then, we use
K-MEANS PLUS algorithm to automatically confirm 5 center points of supply-demand
gap. Through the calculation of similarity, 5 samples are highly
representative and 58 blocks are well distinguished by 5 clusters. Finally,
we use the K-MEANS algorithm to cluster the 288 dimensions model. The 58 blocks are gathered
into 5 clusters and the serial number of each cluster is 0, 1, 2, 3 and 4.

We analyze the similarity of POIs in one cluster and the similarity of POIs between
clusters. Taking into account that the divergence of POIs between clusters is mainly
the divergence in quantity, we choose Euclidean-Distance(E-D) to calculate
the similarity so that It can be intuitively found out the difference
between similarity of POIs in one cluster and similarity of POIs among
clusters. Since cluster 1 and cluster 4 have a few samples, so we
select samples of cluster 0, cluster 2 and cluster 3 to calculate. The
results are shown in the table \ref{tab3-3}:

\begin{table}[htbp]
\centering
\tabcolsep = 5pt
\caption{calculation results of similarity}
\begin{tabular}{|c|c|c|c|c|c|}
\hline
\diagbox{cluster}{E-D}&
0.000001&
0.0000015&
0.000002&
{0.0000025} &
0.000003 \\
\hline
cluster 0&
0&
0.0085&
0.1974&
{0.6131} &
0.8895 \\
\hline
cluster 2&
0&
0&
0&
{0.4667} &
0.83 \\
\hline
cluster 3&
0&
0&
0&
{1} &
1 \\
\hline
Between & \multicolumn{5}{|c|}{~} \\
\hline
cluster 0, 2&
0&
0&
0.1354&
0.4323&
{0.7813}  \\
\hline
cluster 0, 3&
0&
0&
0.0625&
0.4028&
{0.8611}  \\
\hline
cluster 2, 3&
0&
0&
0&
0.25&
{0.75}  \\
\hline
\end{tabular}
\label{tab3-3}
\end{table}

The smaller the E-D certificates the higher the similarity. From table 2, the
numbers in the form represent the proportions within the range of E-D. It
is in appearance that almost all of the similarity of POIs in one cluster is
more higher than the similarity of POIs among clusters, which prove POIs have relationships
with supply-demand gap.

In this Section, we conduct analysis and experiments to demonstrate our assumption - POIs have impact on estimation. However, there are two questions surfaced. One is whether more number of POIs added to XGBoost will be more beneficial to improve estimation accuracy. The other is whether contribution of any POI to estimation is the same. In the next Section, we will probe this two questions.

\subsection{The Different Contributions of POIs}

In this Section, we study the effect of the amount of POIs and the type of POIs on the estimation.

We construct 4 collections of POIs, each collection contains 8 groups. The number of POIs foe each group is 5, 10, 15, 20, 30, 50, 75 and 100. Moreover, the combination types of POI in each group are not the same. We put these groups of POIs into the XGBoost for estimation and use estimation accuracy as evaluation standard. The results are shown in the figure \ref{fig3-1}:

\begin{figure}[!h]
\begin{tabular}{cc}
  \includegraphics[width = 0.48\textwidth]{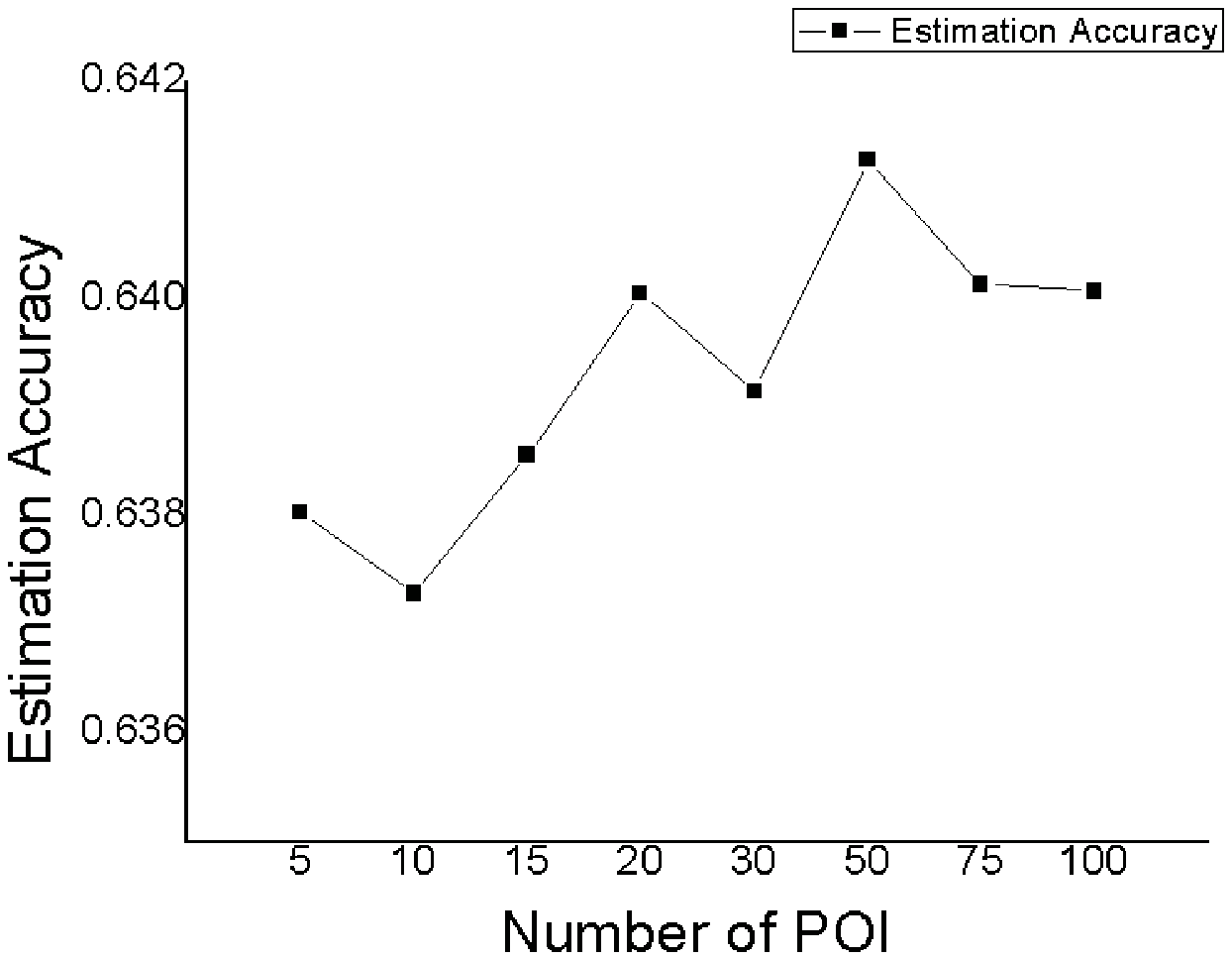}&
  \includegraphics[width = 0.48\textwidth]{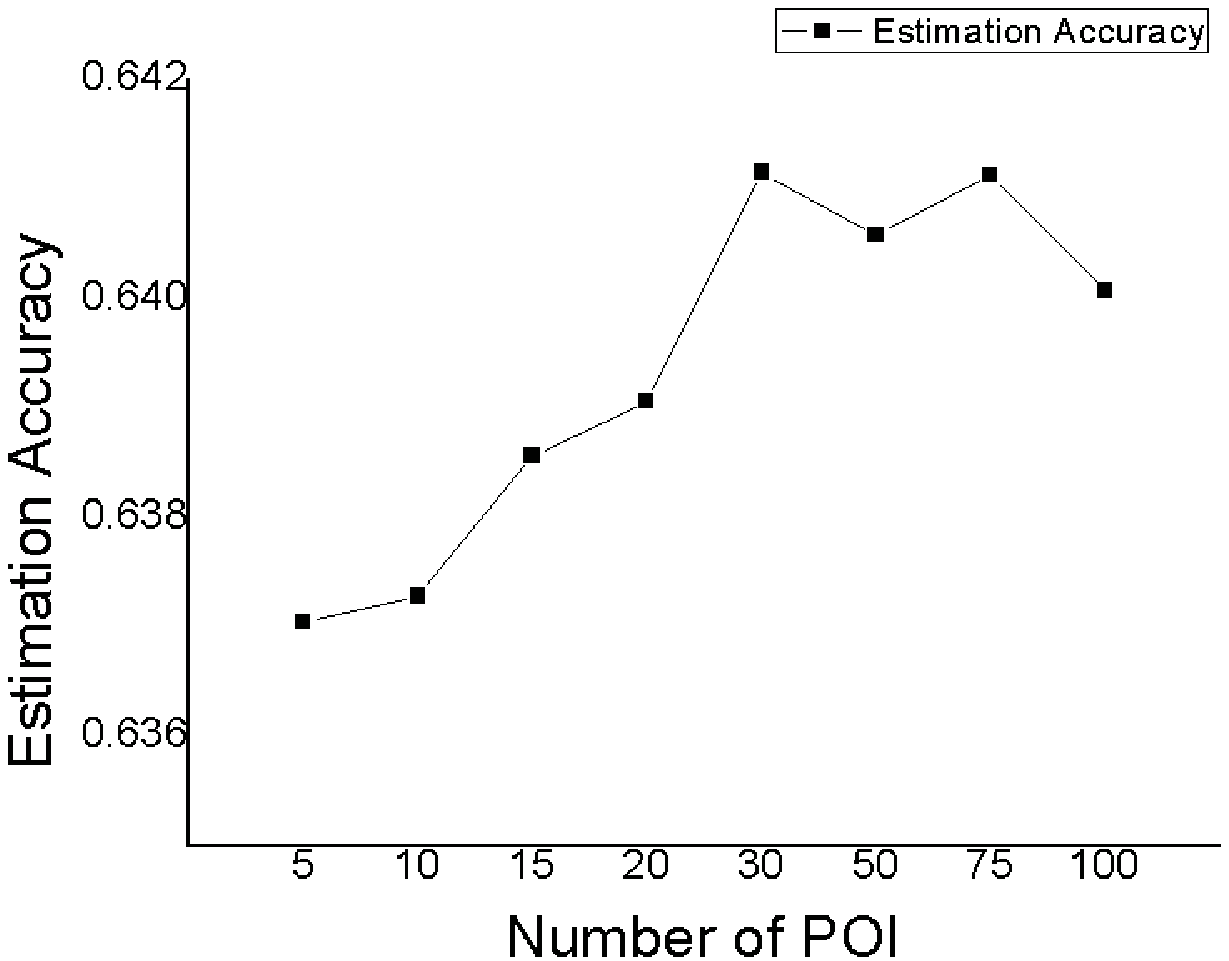}\\
  \qquad (a) &\qquad (b)\\
  \includegraphics[width = 0.48\textwidth]{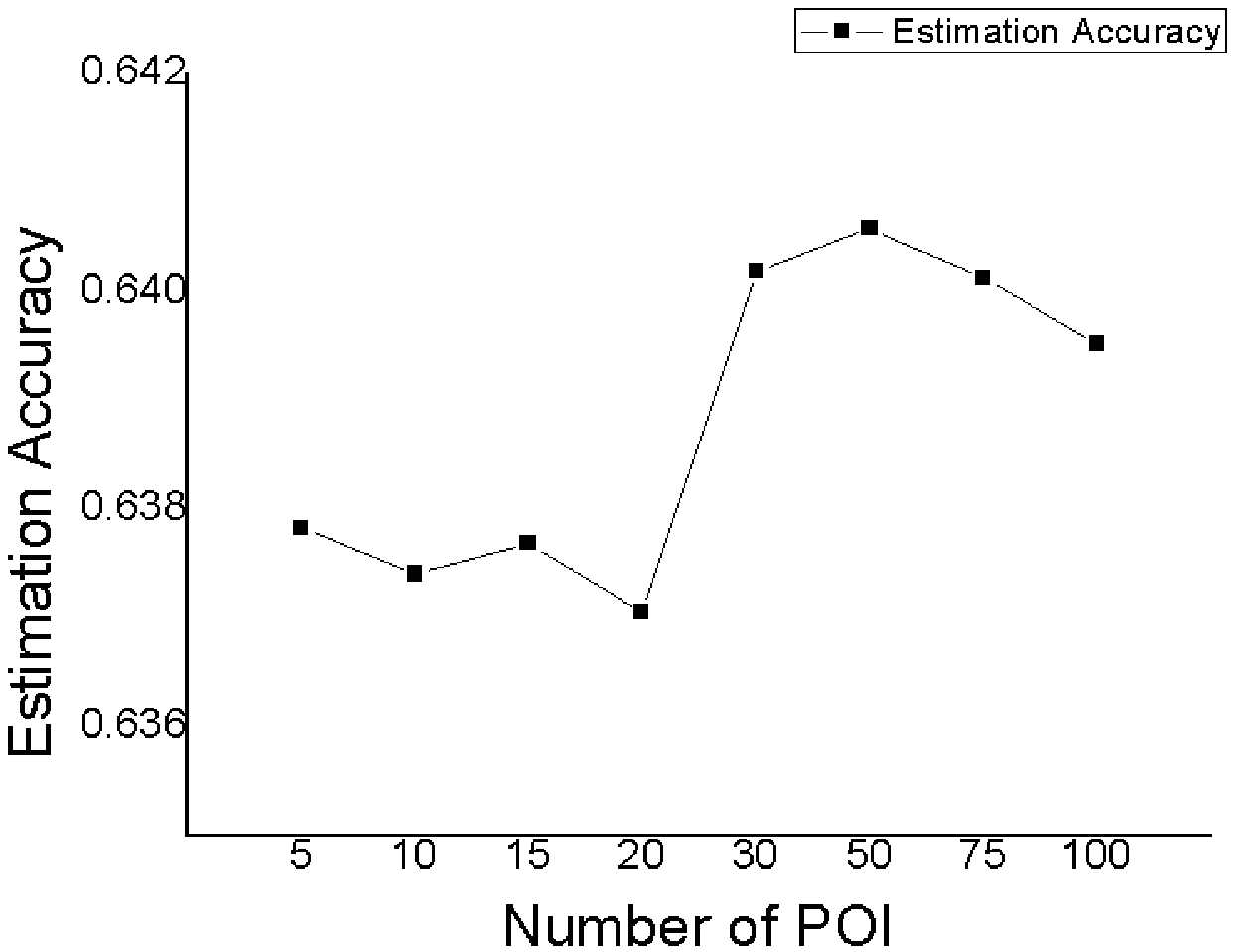}&
  \includegraphics[width = 0.48\textwidth]{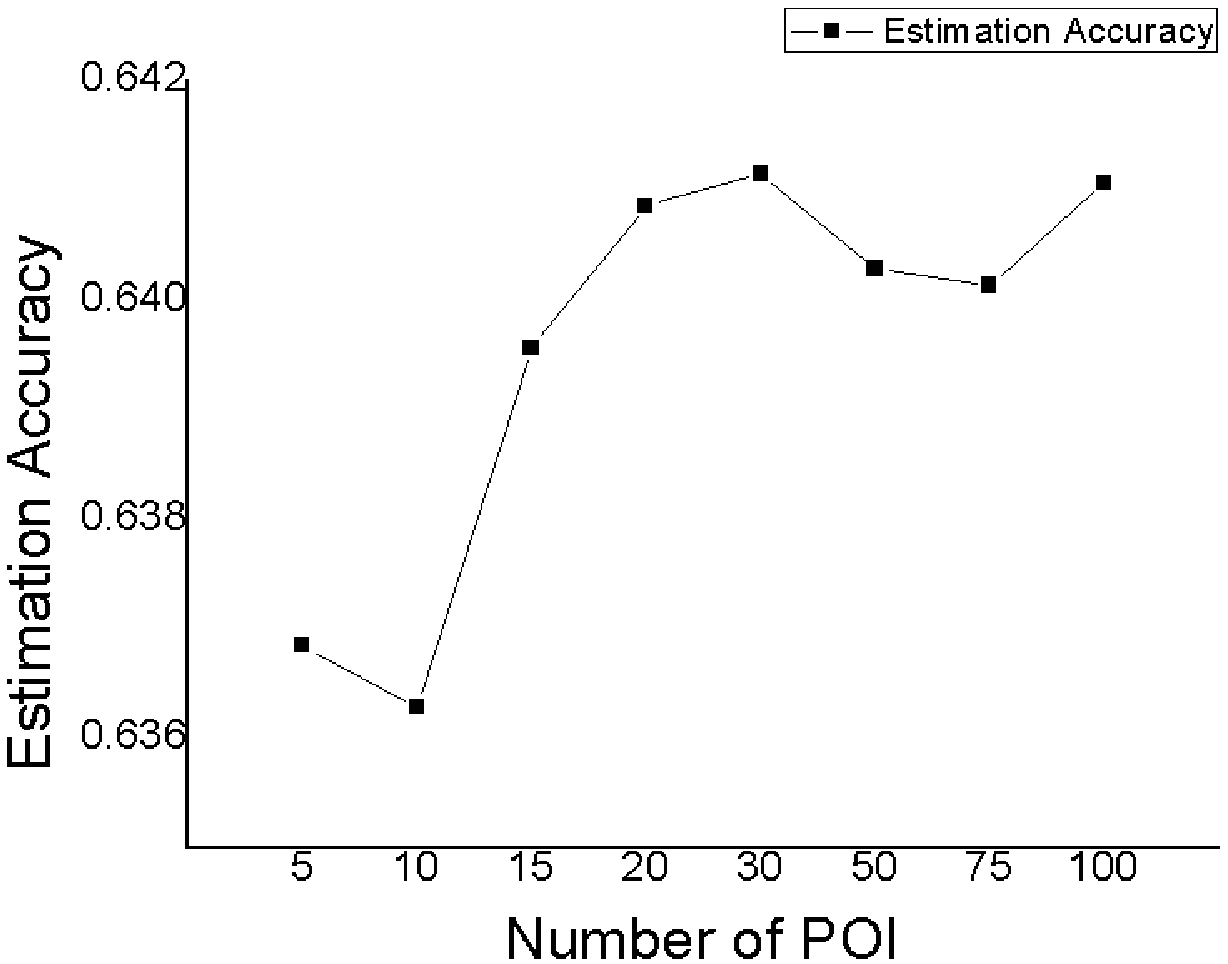}\\
  \qquad (c) &\qquad (d)
  \end{tabular}
  \caption{estimation accuracy of 5, 10, 15, 20, 30, 50, 75, 100 POIs}\label{fig3-1}
\end{figure}

We firstly observe the effect of the amount of POIs on the estimation. From figure \ref{fig3-1}, we observe the overall estimation accuracy rises with the increase of the
amount of POIs. However, more amount of POIs do not mean the more higher estimation accuracy, such as,
the accuracy with 100 POIs is not the highest. Moreover, the XGBoost-SD will take longer time for estimation with more amount of POIs. Thence, the amount of POIs is not a decisive factor. So what kind of impact does the distinction of POIs have on the estimation?

To illustrate affect of different type of POI on estimation, we continue to observe the figure \ref{fig3-1}. From (a), (b), (c), (d), accuracy of the same amount of POIs are distinct, which prove the different categories of POIs have different effects on estimation. The above experiments illustrate the amount and type of POIs can affect the estimation. However, for car-hailing  supply-demand estimation, the amount of POI is not a determinant factor, nor the type of POI. Only combination of amount and type of POIs can better improve the estimation accuracy. Therefore, it is a very important subject that how to select out a few significant POIs.

\subsection{Selection Method}
These above tests illustrate the POIs have influence on predicting gap and the
different amount and type of POIs' effect are distinct. However, how to choose the
important POIs is very challenging. There are 173 categories of POIs
distributed in 58 blocks In each block, the number of POIs and the kinds of
POIs is extremely different. Moreover, all of POIs are represented by
numerical identifier.

In order to effectively extract out POIs, we propose a method called
PPCE-SDG(POI Principal Component Extraction Based on Supply-Demand Gap).
On the basis of clustering blocks, we use PCA algorithm to calculate
absolute value of the eigenvalues of each POI. PCA can transforms the
original data into a set of linearly independent representations by
linear transformation. It can be used to extract the main feature
components of data, which is mainly used for reduction of high dimensional
data. Finally, we extract out significant POIs according with the absolute
value of the eigenvalues of each POI. The PPCE-SDG can be described by pseudo code in algorithm 1:
\begin{algorithm}[!h]
  \caption{PPCE-SDG Process}
  \begin{algorithmic}[1]
  \REQUIRE~~\\
  G: gap data-set\\
P: POI data-set\\
B: the set of the blocks\\
T: the set of the training days

   \ENSURE~~ \\
   POI selected

   \STATE Begin:
\FOR{each block $b_i $ in B}

\FOR{each time slice $t_j $ in T}

\STATE $b_i .$ calculate($\frac{1}{T}\sum\limits {(gap_{i,j} )} )$ //calculate the
average of gap in same time slice of T

\ENDFOR

\ENDFOR

\STATE gap{\_}vector = statistic(B.calculate($\frac{\sum G }{T}))$ //integrate the
mean of gap into a gap vector

\STATE center{\_}points = k-means+(gap{\_}vector) //use k-means algorithm to
confirm center points

\STATE clusters = k-means(center{\_}points, B) //use k-means algorithm to cluster
the blocks

\FOR{each cluster $c_k $ in clusters}

\FOR{each POI $p_m $ in $c_k $}

\STATE $c_k .$ calculate($\frac{1}{B_k }\sum\limits {(p_m )} )$ //calculate the
average of a type POI in each cluster

\ENDFOR

\ENDFOR

\STATE poi{\_}vector = statistic(cluster.calculate($\frac{\sum P }{B}))$ //
integrate the mean of POI into a poi vector

\STATE POI{\_}eigenvalues = PCA(poi{\_}vector) // use PCA algorithm to calculate
the eigenvalues of each type POI

\STATE sort $<$--- Max(POI{\_}eigenvalues)

\STATE return POI{\_}selected
\STATE End
  \end{algorithmic}
\end{algorithm}

PPCE-SDG is strongly logical. It revolves around the data and each step of method has a strict calculation and derivation. Moreover, PPCE-SDG can be combined with XGBoost. In choosing of POI features, PPCE-SDG can replace XGBoost Feature Selection Method(XFSM). When inputting POI features, PPCE-SDG can extract out a few significant POI features and directly put them into XGBoost for prediction. Since the XFSM has to calculate the information gain of each feature, and PPCE-SDG only has low power operation, so the effectiveness of PPCE-SDG is more higher than XFSM in choosing POI features.

\section{Experiment}

\subsection{Evaluation Metric}

\textit{Definition 1 (Supply-demand Gap):} For the d-th day, the supply-demand gap of the time interval [t; t + C) in
area a is defined as the total amount of invalid orders in this time
interval. We fix the constant C to be 10 in this paper and we denote the
corresponding gap as $\text{gap}_\text{a}^{\text{d,t}} $

1) \textit{Error Metrics:} We evaluate the predicted results using the \textit{mean absolute error} (MAE) and the \textit{root mean squared error }(RMSE). Formally, we use $\text{pred}_\text{a}^{\text{d,t}} $to denote the predicted value of $\text{gap}_\text{a}^{\text{d,t}} $. The mean absolute error and the root mean squared error can be computed as follows:
\[
\text{MAE}=\frac{{1}}{{\vert T\vert }}\sum\limits_{(a,d,t)\in T}
{\vert gap_a^{d,t} -pred_a^{d,t} \vert }
\quad
\text{RMSE}=\sqrt {\frac{\text{1}}{{\vert T\vert
}}\sum\limits_{(a,d,t)\in T} {(gap_a^{d,t} -pred_a^{d,t} )^2} }
\]

2) \textit{Estimation accuracy: } Accuracy directly represents the estimation ability. H is the hit number, N is the total number. Accuracy can be computed as follows:
\[
Accuracy = \frac{H}{N}
\]

3) \textit{F1: } Recall Ratio measures the model whether or not can completely gather all elements of test set. Precision Ratio measures the degree of prediction accuracy. F1 is a comprehensive evaluation standard of estimation accuracy. It considers both of Recall Ratio and Precision Ratio. F1 can be computed as follows:
\[
F_1 = \frac{2* Precision}{Precision + Recall}
\]

\subsection{Validation}

In order to verify whether PPCE-SDG is reasonable and helpful to the
prediction, we construct two groups of comparative tests. The first group of experiments demonstrates that PPCE-SDG can extract out a few important POIs. The second group of experiments demonstrates that PPCE-SDG has advantages over the existing approaches.

\subsubsection{4.2.1 Effectiveness}
We put POI data-set into PPCE-SDG and count the contributions of each POI. Afterwards, we divide POIs into three collections according to the contribution degree, which separately mean the POIs of great contribution, POIs of medium
contribution and POIs of small contribution to prediction. In the experiment, the three collections are represented by Max, Med and Min. In each collection, we select POIs of the greatest contribution from 1 to 10 in turn and add them to XGBoost for prediction. We use accuracy as evaluation standards and results are shown in the figure \ref{fig4-1}:
\begin{figure}[!h]
  \centering
  \includegraphics[width = 0.8\textwidth]{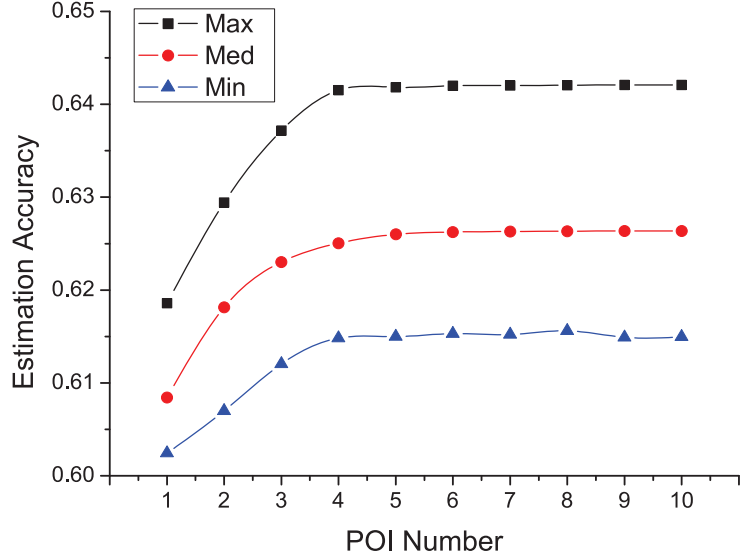}
  \caption{Estimation Results}\label{fig4-1}
\end{figure}

In figure \ref{fig4-1},  we can see that the accuracy of Max is much higher than the Med's and the accuracy of Med is much higher than the Min's, which prove PPCE-SDG can extract out significant POIs. Moreover, when the amount of POIs added to the model reach four, the estimation accuracy is basically stable, which shows that a few POIs play important roles. From the efficiency point of view, Adding fewer POIs to the prediction model means more higher prediction efficiency. The above illustrate PPCE-SDG is effective and helpful for choosing POIs.


\subsubsection{4.2.2 Superiority}

We construct four comparative tests to verify PPCE-SDG has advantages in
extracting out POIs. In Section 4.2.1, we observe that four important POIs can achieve a good estimation results. Therefore, we select the four POIs of the greatest contributions as our experimental group (PPCE-SDG). Then, we choose XGBoost Feature Selection Method (XFSM) and GBDT Feature Selection Method (GFSM) as contral groups. XFSM and GFSM utilizes a gain function to calculate the importance score of features and they works well in feature extraction. The parameters of XFSM and GFSM are fine-tuned through the grid search. We respectively use XFSM and GFSM to select four most important POIs. Moreover, we randomly select four POIs as another contral group (Random). We calculate MAE, RMSE and F1 as evaluation standard. The experimental results are shown in the table \ref{tab4-1}:

\begin{table}[htbp]
\begin{center}
\tabcolsep = 20pt
\caption{performance comparison}
\begin{tabular}{|c|c|c|c|c|c}
\hline
\raisebox{-1.50ex}[0cm][0cm]{Group of POI}&
\multicolumn{3}{|c|}{Evaluation Standard}  \\
\cline{2-4}
 &
MAE&
RMSE&
F1 \\
\hline
XFSM&
3.47&
14.88&
0.67 \\
\hline
GFSM&
3.62&
15.71&
0.66 \\
\hline
PPCE-SDG&
3.41&
14.33&
0.68 \\
\hline
Random&
3.64&
15.82&
0.66 \\
\hline
\end{tabular}
\label{tab4-1}
\end{center}
\end{table}

From Table \ref{tab4-1}, we can see that the Random gap is much larger than the other methods. XFSM provides a much better
prediction result than the GFSM. Rondom is similar to GFSM. XFSM achieves the
best prediction results among groups mentioned above, for both MAE and RMSE.
Our method significantly outperform all the others. Max achieves the best
prediction results for both MAE and RMSE, which demonstrates its prediction
power. The MAE of the XFSM is 1.7{\%} higher than the Max and the RMSE of
the Max is 3.6{\%} lower than the XFSM. F1 of Max significantly outperform
all the others. These indexes demonstrate that Max have better estimation
ability, which prove PPCE-SDG is reasonable and helpful to the prediction.

\section{Conclusion}
In this paper, we study the problem of choosing important POIs for a high
estimation accuracy. At first, we prove the relation between the POIs and
the car-hailing supply-demand gap. Then, we certificate the different categories
of POIs have different effects on estimating.
Moreover, we propose a POIs selection scheme called PPCE-SDG and incorporate it into XGBoost framework to achieve more accurate estimation results. Finally, We conduct
extensive experiment, which show that our method is more effective and powerful in extracting out POIs than the existing methods.

%
\end{document}